\documentclass[fleqn,usenatbib]{mnras}

\usepackage{newtxtext,newtxmath}

\usepackage[T1]{fontenc}

\DeclareRobustCommand{\VAN}[3]{#2}
\let\VANthebibliography\thebibliography
\def\thebibliography{\DeclareRobustCommand{\VAN}[3]{##3}\VANthebibliography}

\usepackage{graphicx}	
\usepackage{amsmath}	

\usepackage{acronym}
\usepackage{siunitx} 

\acrodef{ADC}[ADC]{atmospheric dispersion corrector}
\acrodef{AO}[AO]{adaptive optics}
\acrodef{CCD}{charged-coupled device}
\acrodef{CRED2}[C-RED 2]{First Light C-Red 2 InGaAs detector}
\acrodef{CfAI}[CfAI]{Centre for Advanced Instrumentation}
\acrodef{DM}[DM]{deformable mirror}
\acrodef{ELT}[ELT]{Extremely Large Telescope}
\acrodef{FLAO}{First Light AO}
\acrodef{FWHM}[FWHM]{full width half maximum}
\acrodef{IFU}[IFU]{Integral Field Unit}
\acrodef{KIT}[KIT]{Karlsruhe Institute for Technology}
\acrodef{LBT}[LBT]{Large Binocular Telescope}
\acrodef{LSF}[LSF]{line spread function}
\acrodef{Lili}[Lili]{Little iLocater}
\acrodef{MCF}[MCF]{multi-core fibre}
\acrodef{MCIFU}[MCIFU]{Multi-Core Integral Field Unit}
\acrodef{MFD}[MFD]{mode-field diameter}
\acrodef{MLA}[MLA]{micro-lens array}
\acrodef{MM}[MM]{multi-mode}
\acrodef{MMF}[MMF]{multi-mode fibre}
\acrodef{MOS}[MOS]{multi-object Spectrograph}
\acrodef{MPIA}[MPIA]{Max Planck Institute for Astronomy}
\acrodef{NA}[NA]{numerical aperture}
\acrodef{NCP}[NCP]{non-common path}
\acrodef{NAIR}[NAIR]{Novel Astronomical Instrumentation based on photonic light Reformating}
\acrodef{NIR}[NIR]{near-infrared}
\acrodef{OAP}[OAP]{off-axis parabola}
\acrodef{PL}[PL]{photonic lantern}
\acrodef{PLA}[PLA]{Polylactic Acid}
\acrodef{PSF}[PSF]{point-spread function}
\acrodef{PSD}[PSD]{power spectral density}
\acrodef{QE}[QE]{quantum efficiency}
\acrodef{SM}[SM]{single-mode}
\acrodef{SMF}[SMF]{single-mode fibre}
\acrodef{VLT}[VLT]{Very Large Telescope}
\acrodef{VPH}[VPH]{Volume Phase Holographic}
\acrodef{VPHG}[VPHG]{Volume Phase Holographic Grating}
\acrodef{WFS}[WFS]{wavefront sensor}
\acrodef{ExAO}[ExAO]{extreme adaptive optics}

\title[Lili]{Little iLocater: paving the way for iLocater}

\author[R. J. Harris et al.]{Robert J. Harris,$^{1}$\thanks{E-mail: robert.j.harris@durham.ac.uk}
Jonathan Crass,$^{2,3}$
Marshall C. Johnson,$^{2}$
Andrew Bechter,$^{3}$
Jennifer Power,$^{4}$
\newauthor
Ariadna Calcines Rosario,$^{1}$
Justin R. Crepp,$^{3}$
Eric Bechter,$^{3}$
Brian Sands,$^{5}$
\newauthor
Derek Kopon,  
Steve Ertel,$^{6,7}$
Santiago Barboza $^{8}$
and Andrea Bianco $^{9}$
\\
$^{1}$Centre for Advanced Instrumentation, Department of Physics, Durham University, South Road, Durham, DH1 3LE, UK\\
$^{2}$Department of Astronomy, The Ohio State University, 4055 McPherson Laboratory, 140 West 18$^{\mathrm{th}}$ Ave., Columbus, OH 43210, USA \\
$^{3}$Department of Physics \& Astronomy, University of Notre Dame, 225 Nieuwland Science Hall, Notre Dame, IN 46556, USA\\
$^{4}$Large Binocular Telescope Observatory, 933 N. Cherry Ave, Tucson, AZ 85721-0065, USA \\
$^{5}$Engineering \& Design Core Facility, University of Notre Dame, 626 Flanner Hall, Notre Dame, IN 46556, USA\\
$^{6}$Department of Astronomy and Steward Observatory, The University of Arizona, 933 North Cherry Ave, Tucson, AZ 85721, USA\\
$^{7}$Large Binocular Telescope Observatory, The University of Arizona, 933 North Cherry Ave, Tucson, AZ 85721, USA \\
$^{8}$Max-Planck-Institute for Astronomy, K\"onigstuhl 17, 69117, Heidelberg, Germany \\
$^{9}$INAF - Osservatorio Astronomico di Brera, via E. Bianchi 46, 23807 Merate (LC), Italy 
}

\date{Accepted XXX. Received YYY; in original form ZZZ}

\pubyear{\the\year{}}

\begin{document}
\label{firstpage}
\pagerange{\pageref{firstpage}--\pageref{lastpage}}
\maketitle

\begin{abstract}

Diffraction-limited radial velocity instruments offer a pathway towards improved precision and stability, and the exploration of new parameter spaces at high spatial and spectral resolution. However, achieving the necessary performance requires careful instrument design and considerable on-sky testing. We describe the design and construction of ``Little iLocater'' (Lili), a compact spectrograph that has been used to validate the performance of the front-end fibre-injection system of the iLocater spectrograph. We present the design, assembly, and performance using on-sky data obtained at the Large Binocular Telescope (LBT), including extraction of spectra from standard stars, testing of the atmospheric dispersion corrector to elevations of \qty{40}{\degree}, and spatially resolved spectra from close companion systems. These results show the front-end fibre-injection system is performing as expected and is indicative of iLocater's capabilities once installed at the LBT. 
\end{abstract}




\section{Introduction}\label{sect:intro}  

Spectrographs providing precise radial velocity measurements of stars are a key tool for the detection and characterisation of exoplanets and the stars they orbit. Several next-generation instruments use \ac{AO} for fibre injection (e.g. \citep{Crass:2021, Gibson2019,Lovis2022,Mawet2022,Sliski_2023, Vigan2024}),  allowing increased spatial resolution and spectral resolution, the elimination of spatial modal noise, a stable line spread function, reduction in on-sky background contamination, and a smaller spectrograph to be built, thus improving instrument stability and precision \citep{Bland-Hawthorn2006,Crepp2014}. However, building these instruments remains a challenge, as numerous error sources can combine to reduce measurement precision for the instrument overall \citep{bechter_18,bechter_19_hr4g,bechter_20_pol}. Testing of critical components and observing modes is vital to the ultimate success of the spectrograph. 

\subsection{Fibre Optics in Astronomy}

Fibre optics allow the efficient and flexible transport of light from one location to another and are integrated into many aspects of everyday life. Initially, fibre-fed instruments were designed to couple starlight from seeing limited telescopes. In order to do this efficiently the fibres were required to have cores of order 100 microns in diameter. While large mode field diameters offer high throughput, seeing-limited spectrographs on large telescopes require proportionally larger diffraction gratings to generate sufficient resolution \citep{Bland-Hawthorn2006, Spaleniak2013}. This relationship is particularly problematic in the era of \acp{ELT}, where high resolution seeing-limited spectrographs need to be the size of a small house. 

\subsection{ Single-Mode Fibres and Photonic Lanterns}
\Acp{SMF} are fibres with cores of order \qty{10}{\micro\meter} that only support a single mode (excluding  polarization, which allows two modes to propagate). To couple efficiently into this mode, a near Gaussian beam is required, with a flat phase profile. Interferometric instruments were early adopters of \acp{SMF}, as having only one spatial mode means having to only compensate for one delay (effective index) between telescopes. This has allowed the development of fibre-fed instruments like FLUOR \citep{Scott:2013} at CHARA and VINCI \citep{Kervella2003}, PIONEER \citep{LeBouquin2011} and GRAVITY \citep{gravity:2017} at the \ac{VLT}. 

The major drawback of using a \ac{SMF} input is that unless the image remains diffraction-limited and stable, coupling efficiencies degrade significantly \citep{Jovanovic2017}. Thus, the use of \ac{SMF} in spectroscopy was not considered an effective solution for years. Recently, two different avenues have emerged to make the use of \ac{SMF} in spectroscopy viable. 

The first approach benefits from the increased adoption of \ac{AO} on large-aperture telescopes. \ac{AO} restores the spatial resolving power of a telescope to the diffraction limit, instead of that set by the atmosphere. Background noise is reduced by 1-2 orders of magnitude when focusing light into a smaller region \citep{Crepp2014,Crass2019}  and allows direct injection of the \ac{PSF} into the \ac{SMF} core. The second approach involves using a \ac{PL} to couple a seeing-limited \ac{PSF} from the telescope \citep{Leon-Saval:2005, Birks:2015}. Acting as a large-core fibre (or waveguide array) at the focal plane, the \ac{PL} adiabatically transitions a multi-mode signal input into an equivalent number of \acp{SMF} or waveguides. By transforming a multi-mode input into multiple \ac{SMF} output, the \ac{PL} retains the benefits of throughput and minimizing overall instrument size while offering flexibility in spectral formatting, but at the expense of detector real-estate \citep{Harris2012}. In particular, this feature allows reformatting  of fibres into a pseudo slit, akin to image slicing but using waveguides \citep{Harris:2015,MacLachlan:2017}. This pseduo slit may be used to improve stability \citep{anagnos2018}, improve the utilization of detector pixels, and in certain cases reduce instrument size.

\subsection{The iLocater Spectrograph and Lili}
\label{sec:ilocater}

iLocater is an \ac{AO} and \ac{SMF}-fed, high-resolution, stabilized Doppler spectrograph \citep{Crepp2016} developed for the dual \qty{8.4}{\meter} diameter mirrors of the \ac{LBT} and is due to be commissioned in 2025.  The two telescope beams are received by the Universal Beam Combiner (UBC) of the \ac{LBT} interferometer (LBTI, \cite{Hinz2016,Ertel2020}) located at the common center bent Gregorian focus of the two \ac{LBT} apertures.  The two beams are AO corrected by the telescope's adaptive secondary mirrors and the wavefront sensors of the Single-conjugated adaptive Optics Upgrade for the \ac{LBT} (SOUL) \citep{Pinna2016}.  Once operational, iLocater will receive those beams and inject them into two \acp{SMF}, each corresponding to \qty{\approx 40}{mas} on sky. 
The two \ac{SMF}, along with a calibration fibre, will feed a cross-dispersed Echelle spectrograph that operates in the YJ-bands from \qtyrange{0.97}{1.31}{\micro\meter} \citep{Crass_iLocSpec:2021}. The instrument is designed to deliver high spectral resolving power $R = \lambda / \Delta \lambda$ = \qtyrange{125000}{281000}{} with a median value of \qty{190500}{} and be stabilized to sub-meter-per-second precision. The acquisition camera for the \ac{LBT} SX primary mirror has already been installed and tested using narrow-band filters \citep{Crass:2021}. To test the acquisition camera performance spectroscopically, we have developed \ac{Lili}, a compact miniature spectrograph which was temporarily installed at the \ac{LBT} to further validate the end-to-end system performance and enables the development operational procedures for the observing modes of iLocater.

\subsection{Paper Outline}

In this paper, we describe the first spectral tests of the iLocater front-end fibre-injection system (acquisition camera) using \ac{Lili}, a shoe-box sized spectrographic precursor to the full iLocater instrument. A low resolution spectrograph by design, \ac{Lili} allows validation of the iLocater SX acquisition camera system and \ac{ADC} \citep{Crass:2021} prior to integration and commissioning of the final spectrograph \citep{Crass_iLocSpec:2021}.

In Section \ref{sec:lili}, we describe the \ac{Lili} spectrograph, including optical and mechanical design, calibration, and data reduction. In Section \ref{sec:characterisation}, we describe the as-built lab performance. In Section \ref{sec:on_sky_results}, we describe on-sky performance at the \ac{LBT} for a selection of targets. In Section \ref{sec:discussion}, we discuss what improvements could be made to the spectrograph in the future and in Section \ref{sec:conclusion} we provide concluding remarks. 

\section{Lili}
\label{sec:lili}

To test iLocaters's acquisition camera and fibre-injection system, \ac{Lili} was built to accept a wavelength range similar to the final iLocater bandpass $\lambda$ = \qtyrange{0.97}{1.31}{\micro\meter}. \ac{Lili} was designed to generate a resolving power of $R \approx 1500$, allowing resolution of spectral lines, and to run at speeds of \qty{100}{\hertz}, allowing for real-time analysis of effects such as wavelength-dependent throughput. 

To minimize costs for the spectrograph, components from other projects were re-used as the basis for system design and were augmented with COTS components when needed. Three key elements drive the design: the \ac{SMF} used by the iLocater team, the triple stacked \ac{VPH} grating from the \ac{MCIFU}, and the \ac{CRED2} detector.  Table \ref{tab:CRED2} shows the parameters of the \ac{CRED2} detector, which has 640x512 pixels of \qty{15}{\micro\meter}.

\begin{table}
	\centering
	\caption{ Parameters of the \ac{CRED2} detector as per specifications. Note, we only specify the parameters for low and high gain as these were the only modes used on-sky.}
	\label{tab:CRED2}
	\begin{tabular}{cc}
		\hline
        Parameter & Value \\
        \hline
        Pixels & 640x512 \\
        Pixel pitch (\qty{}{\micro\meter}) & 15 \\
        Max frame rate (Hz) & 400 \\
        Read noise (e-) & <30  \\
        Dark current (e-/p/s)  & <600 \\
        Full well capacity (ke-) & 1400 (low) 33 (high) \\
		\hline
	\end{tabular}
\end{table}

\subsection{Optical Design}
\label{sec:optical_design}

\begin{figure}
    \centering
    \begin{tabular}{c} 
    \includegraphics[trim={0 5cm 0 0},clip, width=\columnwidth]{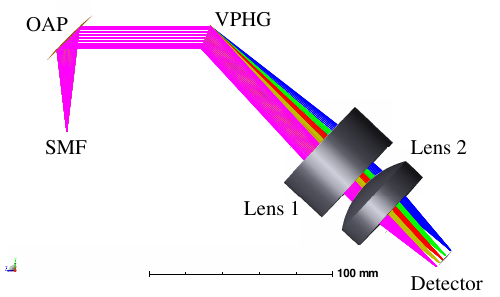}
    \end{tabular}
    \caption{Shaded rendering of the optical design of the \ac{Lili} spectrograph, with the light path coloured by wavelength. From left to right, the light from the \ac{SMF}  is collimated by an \ac{OAP}, which feeds a \ac{VPH} grating. Light is then refocused onto the detector by a TTL200-S8 and AC-508-080-C lens.}
    \label{fig:Lili_optical_design}
\end{figure}

The optical design of \ac{Lili} is shown in Fig. \ref{fig:Lili_optical_design}. A \ac{SMF} injects broadband light into the system. The output of the \ac{SMF} is modeled as a Gaussian beam, with a $1/e^2$ \ac{NA} = 0.12, which corresponds to the fibre supplied by the iLocater team. As the \ac{SMF} acts as an on-axis point source, an \acl{OAP} (MPD129-P02, Thorlabs)  is used to provide a collimated beam. The collimated light feeds a spare triple stacked \ac{VPHG} from the \ac{MCIFU} spectrograph \citep{Haffert2020}.  The grating, which is set at \qty{21}{\degree} to the beam, is slightly off the design angle of \qty{22}{\degree}, reducing efficiency by approximately \qty{8}{\percent}. The grating is composed of three gratings, nominally rotated at \qty{3}{\degree} with respect to each other, allowing separation of three orders on the detector. The resulting diffraction pattern is then re-imaged using two lenses, a TTL200-S8 (Lens 1 in Fig. \ref{fig:Lili_optical_design}) and AC-508-080-C (Lens 2 in  Fig. \ref{fig:Lili_optical_design}), both from Thorlabs. Unlike the \ac{MCIFU}, wavelengths longer than approximately \qty{1.31}{\micro\meter} are not transmitted through the iLocater fibre injection system, thus only two of the three \ac{VPH} gratings are used. It should be noted that the grating was designed to eventually be used with a HAWAII-2 detector and in this system we reduce the image size by a factor of approximately four to fit onto the smaller \ac{CRED2}. This reduced the spectral resolving power of the system because of undersampling of the \ac{LSF}, which is discussed further in Section \ref{sec:on_sky_results}. 

\subsubsection{Lili Resolving Power}

\begin{figure}
    \centering
    \begin{tabular}{c} 
    \includegraphics[trim={0cm 4.5cm 0cm 0cm},width=\columnwidth]{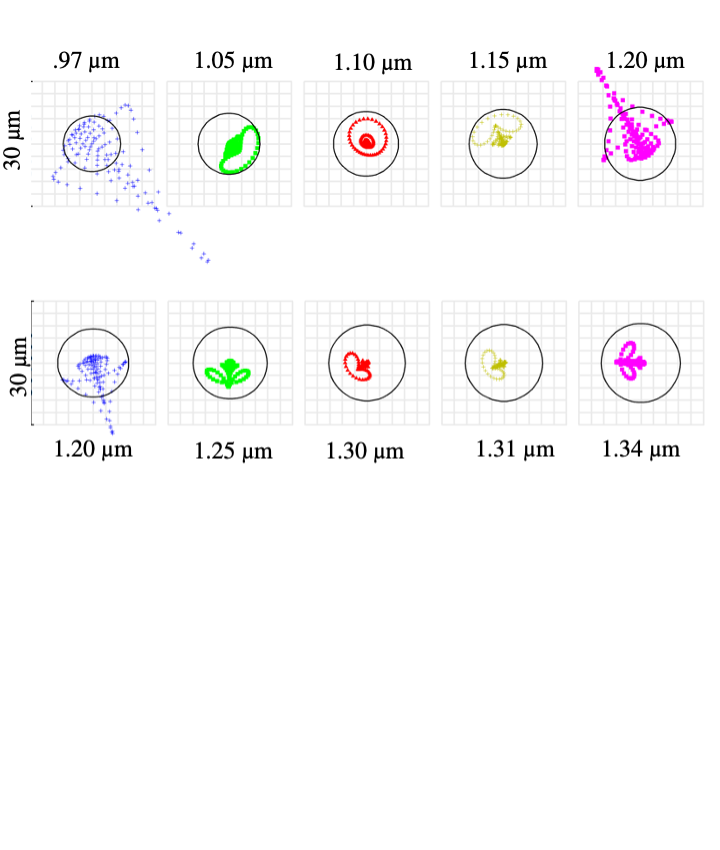}
    \end{tabular}
    \caption{Spot diagrams for both orders of the Lili spectrograph. The first order runs from \qtyrange{.97}{1.2}{\micro\meter} and the second from \qtyrange{1.2}{1.34}{\micro\meter}. In practice the spectrograph is limited by the iLocater fibre link, which accepts \qtyrange{.97}{1.31}{\micro\meter}. The spot diagrams are scaled to \qty{30}{\micro\meter}, which corresponds to two pixels on the \ac{CRED2}. For the majority of both orders the light is confined within a two by two square of pixels.}
    \label{fig:Lili_spots}
\end{figure}

Fig. \ref{fig:Lili_spots} shows the resulting spot diagram on the detector for wavelengths of the two orders incident on the \ac{CRED2}. With the exception of the edges of each order, all of the resulting spots are smaller than the Airy radius. In physical units, the images spots are less than \qty{30}{\micro\meter} in diameter corresponding to two pixels on the \ac{CRED2}. From the Nyquist-Shannon theorem, \ac{Lili} is undersampled on the detector and in the ideal case of perfect alignment the resolving power will be limited by the pixel size, not image quality.

\begin{figure}
    \centering
    \begin{tabular}{c} 
    \includegraphics[trim={0cm 2.5cm 0cm 0cm},clip,width=\columnwidth]{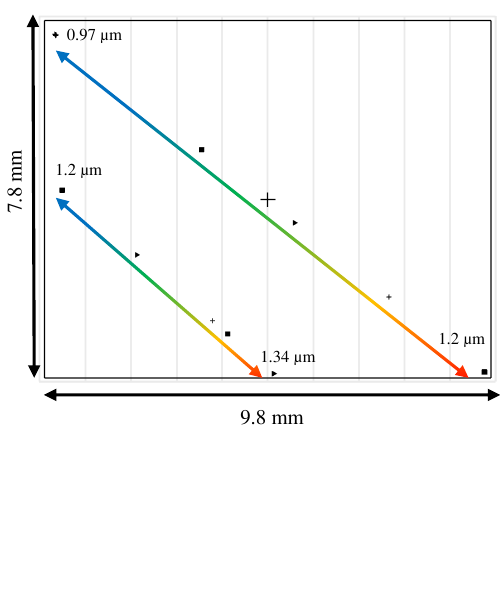} \\
    \end{tabular}
    \caption{Footprint of the two orders imaged onto the detector. The first order (\qtyrange{0.97}{1.2}{\micro\meter}) crosses the full detector area diagonally, while the second sits below (\qtyrange{1.2}{1.34}{\micro\meter}). Note that there is a difference in slope between the first and second orders due to the rotation of the diffraction gratings.}
    \label{fig:Lili_footprint}
\end{figure}

\begin{figure}
    \centering
    \begin{tabular}{c} 
    \includegraphics[width=\columnwidth]{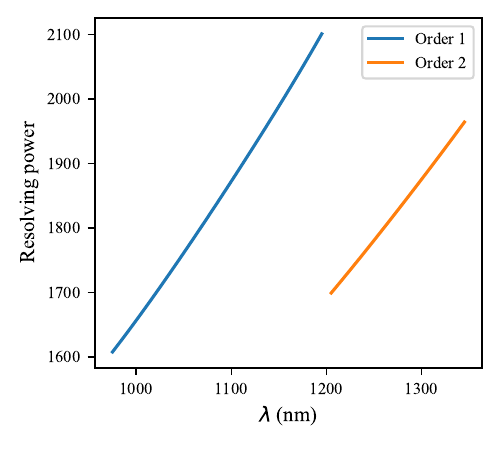}
    \end{tabular}
    \caption{Theoretical resolving power versus wavelength for order one and order two of \ac{Lili}. This optical design meets the required resolving power of $R=1500$.}
    \label{fig:Lili_res}
\end{figure}

In order to maximize resolving power, while also maximizing wavelength coverage, the \ac{CRED2} is rotated by a slight angle, placing the longest order (order one) diagonally from corner to corner (see Fig. \ref{fig:Lili_footprint}). By doing so, this gives a theoretical resolution of $\approx$ \qtyrange{1600}{2100}{} for order one and $\approx$ \qtyrange{1700}{2000}{} for order two (see Fig. \ref{fig:Lili_res}).

\subsubsection{Theoretical spectrograph transmission}

\begin{figure}
    \centering
    \begin{tabular}{c} 
    \includegraphics[width=\columnwidth]{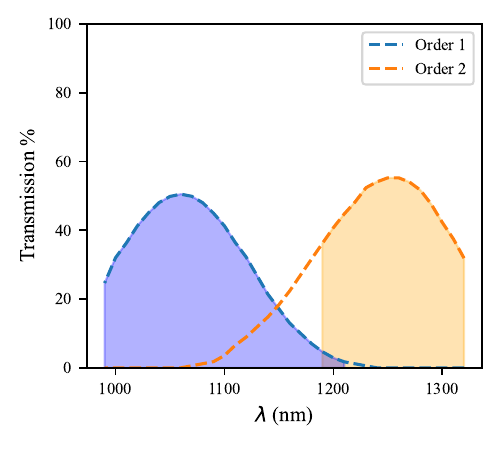}
    \end{tabular}
    \caption{Estimated spectrograph throughput based on commercially available values for lenses and theoretical values for the \ac{MCIFU} gratings. Individual efficiencies of the two spectrograph orders are plotted as dotted lines. The filled area underneath indicates the parts of the spectrum from each order incident on the \ac{CRED2} detector.}
    \label{fig:Lili_transmission}
\end{figure}

The spectrograph transmission was estimated using commercially available values for the lenses and \ac{OAP} from Thorlabs, the \ac{QE} numbers for the \ac{CRED2} from First Light and the efficiency numbers for the gratings measured by the \ac{MCIFU} team, including a factor to account for the grating being used off of its design wavelength. As the Zemax model showed no vignetting in the optical design, no additional factors were added. The transmission across the Lili wavelength range is shown in Fig. \ref{fig:Lili_transmission}. A peak transmission of $\approx$50\% is reached in each order, consistent with the \ac{MCIFU}.

\subsubsection{Optical sensitivity}
\label{sec:optical_tolerancing}

A sensitivity study of the optical components was performed in ANYSYS Zemax to evaluate the feasibility of the \ac{Lili} design. Alignment of individual elements was allowed to vary by \qty{0.2}{\micro\meter} and \qty{0.2}{\degree}, with focus adjustment of the detector acting as a free parameter to compensate and improve image quality. The energy contained in one pixel for wavelengths in the first order (\qtyrange{0.97}{1.2}{\micro\meter}) was used as a metric for evaluation. The most sensitive element was found to be the \ac{OAP}, with $x$ and $y$ centering reducing energy by 50\%; tip / tilt of the \ac{OAP} contributed $\approx$ 1\% reduction in enclosed energy. All other elements in the spectrograph had less than 1\% influence on the amount of light contained within one pixel and were considered negligible.

\subsection{Mechanical Design} 
\label{sec:mechanical design}

\begin{figure}
    \centering
    \begin{tabular}{c} 
    \includegraphics[trim={0 5cm 0 0},clip, width=\columnwidth]{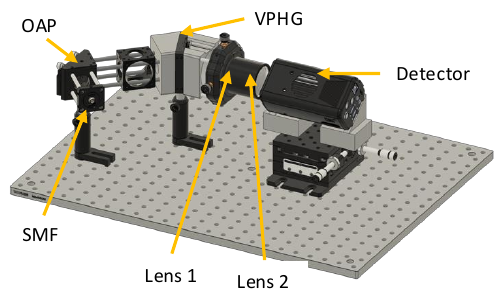}
    \end{tabular}
    \caption{CAD rendering of the \ac{Lili} spectrograph. On the left the iLocater \ac{SMF} that takes light from the acquisition camera injects light into \ac{Lili}. The light is collimated by the OAP. The cage plates and 3D printed parts ensure the triple stacked \ac{VPH} grating is correctly angled for dispersion. The dispersed light is finally focused by two lenses onto the \ac{CRED2} detector.}
    \label{fig:Lili_mech_design}
\end{figure}

A CAD rendering of the \ac{Lili} spectrograph is shown in Fig. \ref{fig:Lili_mech_design}. The system relies on cage mounts and rods for mechanical alignment and rigidity. As section \ref{sec:optical_tolerancing} already demonstrated, the \ac{OAP} alignment is critical and thus the \ac{OAP} was mounted in a fine adjustment mount (Thorlabs KCB1P/M). As with the \ac{MCIFU}, two holders were printed in \ac{PLA} to hold the \ac{VPH} grating at the correct angle. Experience with the \ac{MCIFU} had shown that having accessible rotation adjustment for the \ac{VPH} gratings simplified alignment, so a custom rotation adapter was manufactured and placed inside a LCRM2A/M from Thorlabs. For ease of co-alignment, the two lenses are placed in a lens tube and mounted on an $x$-$y$ stage (CXY2A, Thorlabs). The \ac{CRED2} is liquid cooled to reduce dark current is also placed on an $x$-$y$ stage to allow focus and horizontal position optimization. 

\subsection{Calibration and Data Reduction}
\label{sec:calibration_reduction}

\begin{figure}
    \centering
    \begin{tabular}{c} 
    \includegraphics[width=\columnwidth]{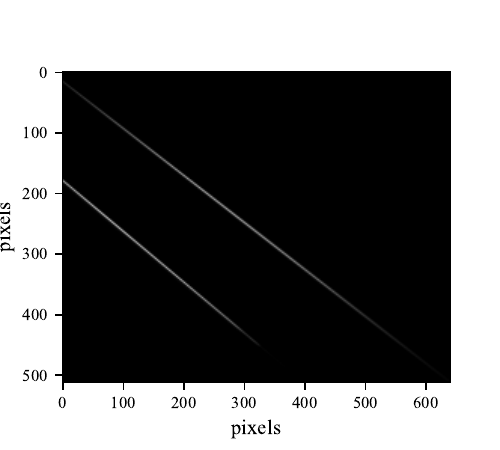}
    \end{tabular}
    \caption{Example raw image of a Halogen spectra taken with the \ac{CRED2} detector. The two orders are visible: the first order runs across the center of the image from top left to bottom right (\qtyrange{0.97}{1.2}{\micro\meter}) and the second order is on the bottom left (\qtyrange{1.2}{1.34}{\micro\meter}). Similar images were used to establish the spectral trace for processing.}
    \label{fig:halogen_raw}
\end{figure}

Several different light sources were used for alignment and calibration. Initial alignment was performed using a \qty{1060}{\nano\meter} laser (S1FC1060, Thorlabs), and a Halogen lamp (OSL2IR, Thorlabs) was used to align the spectrograph orders to the detector (see Fig. \ref{fig:halogen_raw} for an example). A super-continuum source (WhiteLase Micro, Fianium) was used with narrow-band filters (FKBIR kit, Thorlabs) to ensure that the wavelength range was correctly covered. A filter kit was also used to measure transmission across the desired wavelength range (see Section \ref{sec:measured_transmission}). The fully assembled spectrograph was flat-fielded using the halogen lamp and wavelength calibration was performed using an Argon pencil lamp (6030, Newport).

We extracted spectra using standard processing steps, with a custom Python implementation. Extraction proceeded separately for each of the two spectral orders, which were combined into a single FITS file at the end. The spectral orders have negligible curvature but the dispersion direction is rotated at an angle of \qty{38}{\degree} for order one and \qty{40}{\degree} for order two with respect to the rows of the detector. This is \qty{1}{\degree} less than the nominal design in rotation (\qty{3}{degrees}, see Section \ref{sec:optical_design}). As the spectral orders are well separated this has no impact. We rectified this rotation with a simple rotational interpolation to align the dispersion with the columns. We used the optimal extraction algorithm \citep{Horne:1986} to produce a 1d spectrum, using the Halogen lamp spectrum to compute the spectral profile necessary to extract the science spectra. This profile was used to perform optimal extraction of science spectra.

We computed the wavelength solution in two steps. First, we used spectra obtained through the series of narrow-pass filters to create a rough wavelength solution. Second, we extracted a spectrum from the Argon lamp. We identified the Ar peaks, and used the rough wavelength solution to match the peaks to an Ar line list sourced from NIST; several lamp lines not present in the Ar line list were discarded by hand. A quadratic fit to the expected wavelengths versus measured pixel positions of the lines produced a wavelength solution with a dispersion of \qty{0.03}{\nano\meter} in order one and \qty{0.05}{\nano\meter} in order two.

We removed the blaze function and normalized the continuum in a two-step process. First, we divided the spectra by the Halogen flat lamp spectra. This mostly removed the blaze function, as well as a component of high-frequency noise due to striping in the \ac{CRED2}. As the shape of the flat lamp spectrum is not the same as the observed spectra, there is still a low-order shape to the blaze-corrected spectrum. We removed the low order shape by performing an iterative spline fit to the data, rejecting points well below and above the continuum in order to remove absorption lines and isolate the continuum.

 Finally, we performed a flux calibration on the extracted spectrum using Vega as a reference. We built a synthetic model from Caloec \citep{Bohlin_2014} convolved with the ESO SKYCALC atmospheric transmission model \citep{Noll2012,Jones2013,Moehler2014} and interpolated it to the data. The data is divided by the model to get the ratio of flux. We smooth this ratio using a 3rd order Savitzky-Golay filter, with a window length of 200 pixels. The data is then multiplied by the flux ratio to retrieve the flux calibrated spectrum.

\section{On-sky testing at the Large Binocular Telescope}
\label{sec:characterisation}

Assembly, characterization, and on-sky testing of \ac{Lili} with the iLocater acquisition camera took place in May 2024. We were allocated two half nights for on sky tests, which required median-seeing conditions or better to allow efficient \ac{AO} system operation. 

\begin{figure}
    \centering
    \begin{tabular}{c} 
    \includegraphics[width=\columnwidth]{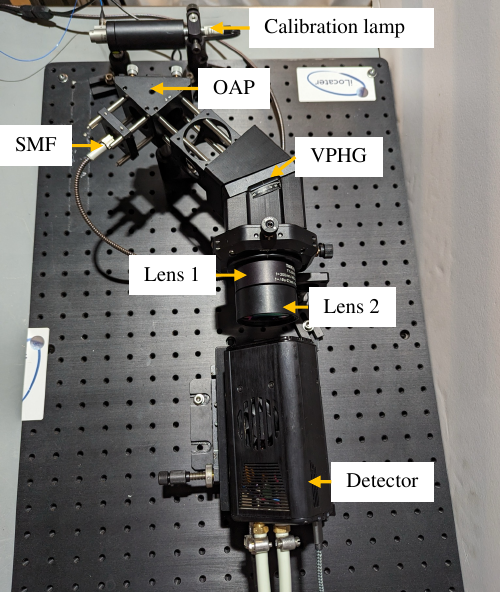}
    \end{tabular}
    \caption{The assembled \ac{Lili} spectrograph and a calibration lamp on a breadboard at the \ac{LBT}. At the top of the image an Argon calibration lamp is placed inside a lens tube. An \ac{OAP} focuses the light into a fibre which is fed through the iLocater system back to the spectrograph. At the bottom of the image (see Fig \ref{fig:Lili_mech_design} for reference), the iLocater \ac{SMF} that takes light from the acquisition camera injects light into \ac{Lili}. The light is collimated by the OAP. The cage plates and 3D printed parts ensure the triple stacked \ac{VPH} grating is at the correct angle for dispersion. The dispersed light is focused by two lenses onto the \ac{CRED2} detector.}
    \label{fig:Lili_as_built}
\end{figure}

\ac{Lili} was assembled inside the telescope pier on level 3L, where iLocater will eventually be located. The fibres that will feed iLocater were already installed. The \ac{CRED2} was attached to the \ac{LBT} chilled water system and was cooled to \qty{-40}{\celsius} for all of the tests. An annotated image of the as-built spectrograph can be seen in Fig. \ref{fig:Lili_as_built}. The spectrograph is built on an aluminium breadboard, with vibration damping feet (AV5, Thorlabs) and placed in an enclosure to minimize stray light entering the system. Temperature control and vibration reduction measures that will be used with iLocater were not in place during this visit. To minimize any potential impact from people changing the temperature or causing vibrations, \ac{Lili} was operated remotely from the telescope control room. 

\subsection{Lili Measured Transmission}
\label{sec:measured_transmission}

The transmission of the spectrograph was measured by illuminating the input fibre with the iLocater supercontinuum source. The light was filtered using \qty{10}{\nano\meter} bandpass filters (FKBIR kit, Thorlabs) located in the collimated beam before the \ac{VPH} grating. The power from the fibre input to the spectrograph and the power at the detector was measured using two separate power meter sensors (S120C and PM16-122, Thorlabs) to cover the whole bandpass. The ratio of the output power against the input power was then taken, giving the transmission of the spectrograph. It should be noted that the measured power was a combination of both spectral orders. 

\begin{figure}
    \centering
    \begin{tabular}{c} 
    \includegraphics[width=\columnwidth]{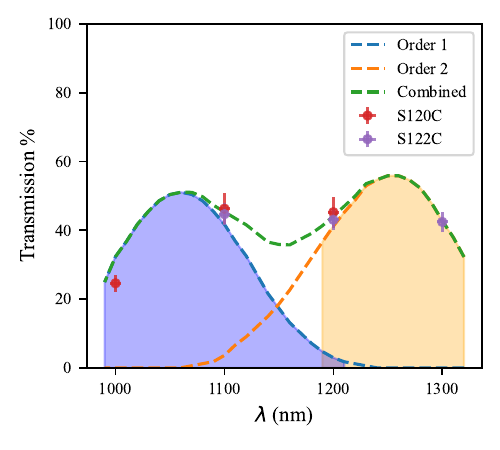}
    \end{tabular}
    \caption{The transmission of \ac{Lili} at the \ac{LBT} vs the theoretical transmission (see Fig. \ref{fig:Lili_transmission}). The values are measured using filters with a \qty{10}{\nano\meter} \ac{FWHM}, supercontinuum source and power meter. Note that due to the large power meter area we measure the combined power from both orders simultaneously (the combined theoretical line). The  difference ($\approx$ \qty{3}{\percent}) between the two measurements at \qty{1200}{\nano\meter} is ascribed to the low efficiency of the S120C power meter at this wavelength and is within the quoted error from the vendor.}
    \label{fig:LBT_transmission}
\end{figure}

Results are shown in Fig. \ref{fig:LBT_transmission} and show very good agreement with the expected transmission from Fig. \ref{fig:Lili_transmission}. There is a discrepancy of up to 3\% between the transmission measured by the two power meters at the overlap point; this is ascribed to the sensor measurement accuracy at these wavelengths, which is quoted as \qty{7}{\percent} and \qty{5}{\percent} from the vendor specifications.

\subsection{Order size and measured resolving power}

 To estimate the magnification of the system we use the known Gaussian $1/e^{2}$ \ac{MFD} values from \cite{Crass:2021}. For order one the supplied fibre has a $1/e^{2}$ \ac{MFD} of \qty{5.8}{\micro\meter} 
at \qty{980}{\nano\meter}, increasing to \qty{\approx 7}{\micro\meter} by \qty{1200}{\nano\meter}. For order 2 this is \qty{\approx 7}{\micro\meter} at \qty{1200}{\nano\meter} increasing to \qty{\approx 7.8}{\micro\meter} at \qty{1310}{\nano\meter}. To calculate the output, we fitted a Gaussian to the spectrally flattened \ac{LSF} of each order and calculate the average $1/e^{2}$ value of each order. This is calculated as \qty{43}{\micro\meter} in order one and \qty{34}{\micro\meter} in order two. We estimate the magnification as 6.14-7.4 for order one and 4.4-4.8 in order two. This indicates that order one was slightly defocused when compared to the design, likely due to a tilt on the detector stage.

\begin{figure}
    \centering
    \begin{tabular}{c} 
    \includegraphics[width=\columnwidth]{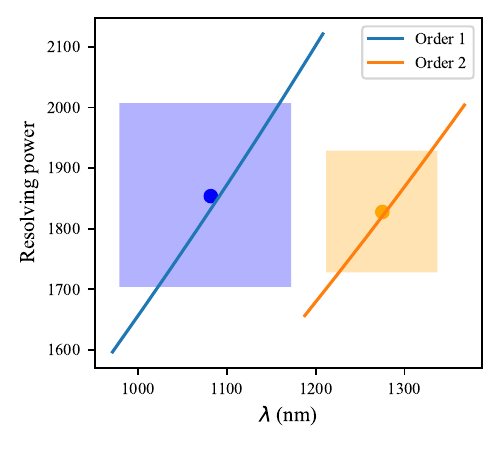}
    \end{tabular}
    \caption{Calculated resolving power of the as-built Lili spectrograph, with order one in blue and order two in orange. The shaded box shows the 1-$\sigma$ resolving power estimates from the Argon emission lines within each order. The points within the box show the median resolving power of that order. We also overplot the 
    resolving power from the design values, showing good agreement.}
    \label{fig:Lili_resolving power}
\end{figure}

The resolving power of the as-built spectrograph was estimated using the spectra of the Argon lamp. We used the calculated parameters of the emission lines, converting the standard deviation of the fitted Gaussian to the \ac{FWHM}; this number was rounded up to an integer number of pixels. The resolving power was then estimated using $R = \lambda / \delta\lambda$. 
The results are shown  in Fig. \ref{fig:Lili_resolving power}, with order one shown in blue and order two shown in orange.  The shaded box shows the 1-$\sigma$ resolving power estimates from the Argon emission lines and the points within the box show the median resolving power of that order. The measured resolving power shows good agreement with the design. 

\section{Testing On Sky}
\label{sec:on_sky_results}

On-sky testing at the \ac{LBT} occurred between 0:00-05:30 and 18:30-24:00 local time (UT-7) on the 17th of May 2024 in two half night blocks. A selection of the observed targets are listed in Table \ref{tab:targets}. \emph{Vega} and \emph{Arcturus} were chosen as they were both high in the sky and already well-characterized by other instruments; they provided validation for the spectrograph measurements. \emph{27 Her} was chosen to test the \ac{ADC} performance given its brightness and low elevation. Finally, \emph{41 Vir} is a known close-companion system and at the time of observation was estimated to have a separation of $\approx$ \qty{50}{mas}; it was chosen to test the ability of the iLocater SX acquisition system to isolate and record individual spectra of close companions. 

\begin{table}
	\centering
	\caption{Table of observations, conditions, and detector parameters for Lili observations. Target  magnitudes are taken from Simbad. The natural seeing is measured using the DIMM at the \ac{LBT} using a bright star within \qty{25}{\degree} of the target star. The frame rate and gain are for the \ac{CRED2}. The test column lists hardware that was being tested.}
	\label{tab:targets}
	\begin{tabular}{cccccc}
		\hline
		 Target & $J$  & DIMM Seeing & Frame rate & Gain & Test\\
       & (mag) & ($\arcsec$) & (Hz) &  & \\
		\hline
		 Vega  & -0.177  & 0.7-1.4 & 100 & low & Spectrograph \\
		  Arcturus & -2.252 & 0.8-1.2 & 100 & low & Spectrograph \\
		 27 Her & 0.858 & 1.3-2.5 & 100 & high & ADC \\
  		 41 Vir & 5.7 & 0.8-1.2 & 2 & high & AO \\
		\hline
	\end{tabular}
\end{table}

\subsection{Spectral extraction: Vega and Arcturus}

Observations of \emph{Vega} and \emph{Arcturus} allowed the \ac{CRED2} to run at \qty{100}{\hertz} at low gain given the brightness of the targets. The reduced spectra were used to provide calibration for \ac{Lili} and test the data reduction pipeline.

\begin{figure*}
    \centering
    \begin{tabular}{c} 
    \includegraphics[width=\textwidth]{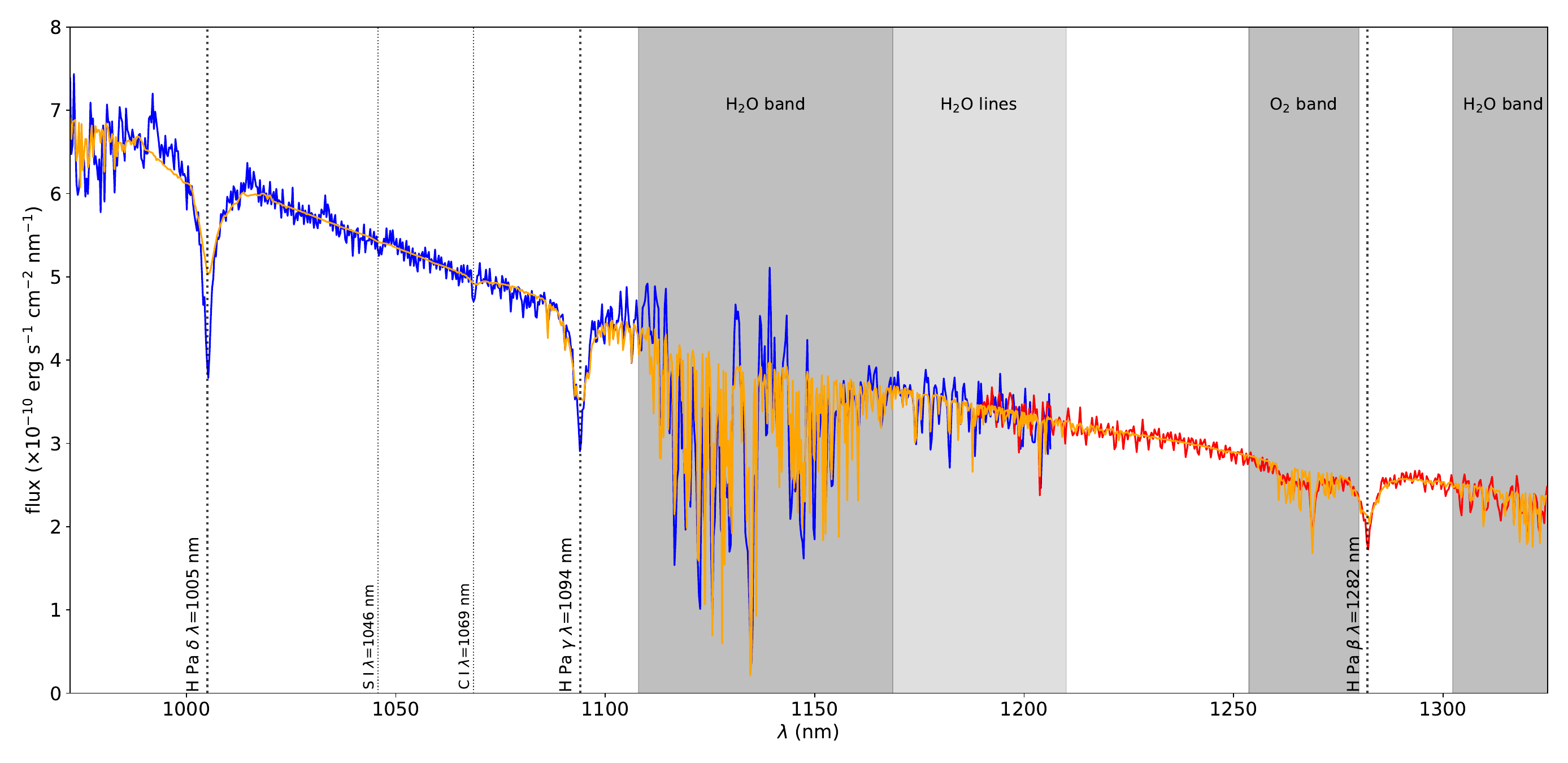}
    \end{tabular}
    \caption{Spectrum of \emph{Vega} recovered by the \ac{Lili} spectrograph over-plotted with a synthetic model. Order one is shown in blue, order two is shown in red, and the synthetic model is shown in orange. Stellar and telluric features are annotated on the plot. Overall, \ac{Lili} recovers the spectral features of $Vega$ in both orders.  }
    \label{fig:Lili_Vega}
\end{figure*}

The extracted spectra of \emph{Vega} is shown in Fig. \ref{fig:Lili_Vega}. The data were reduced as described in Section \ref{sec:calibration_reduction}. We were able to identify several expected lines (H Pa $\delta$, H Pa $\gamma$, H Pa $\beta$), as well as potential metal lines from Si~\textsc{i} and C~\textsc{i}. 

 To estimate the flux coming from iLocater acquisition camera performance we chose \qty{1050}{\nano\meter} as our zero point. Integrating across the spectra over a \qty{1}{\nano\meter} provides approximately \qty{1000}{ADU}, which at low gain converts to \qty{8.5e5}{e^{-}} or \qty{1.2e7}{phs^{-1}\nano\meter^{-1}} at the detector.

\begin{figure*}
    \centering
    \includegraphics[width=\textwidth]{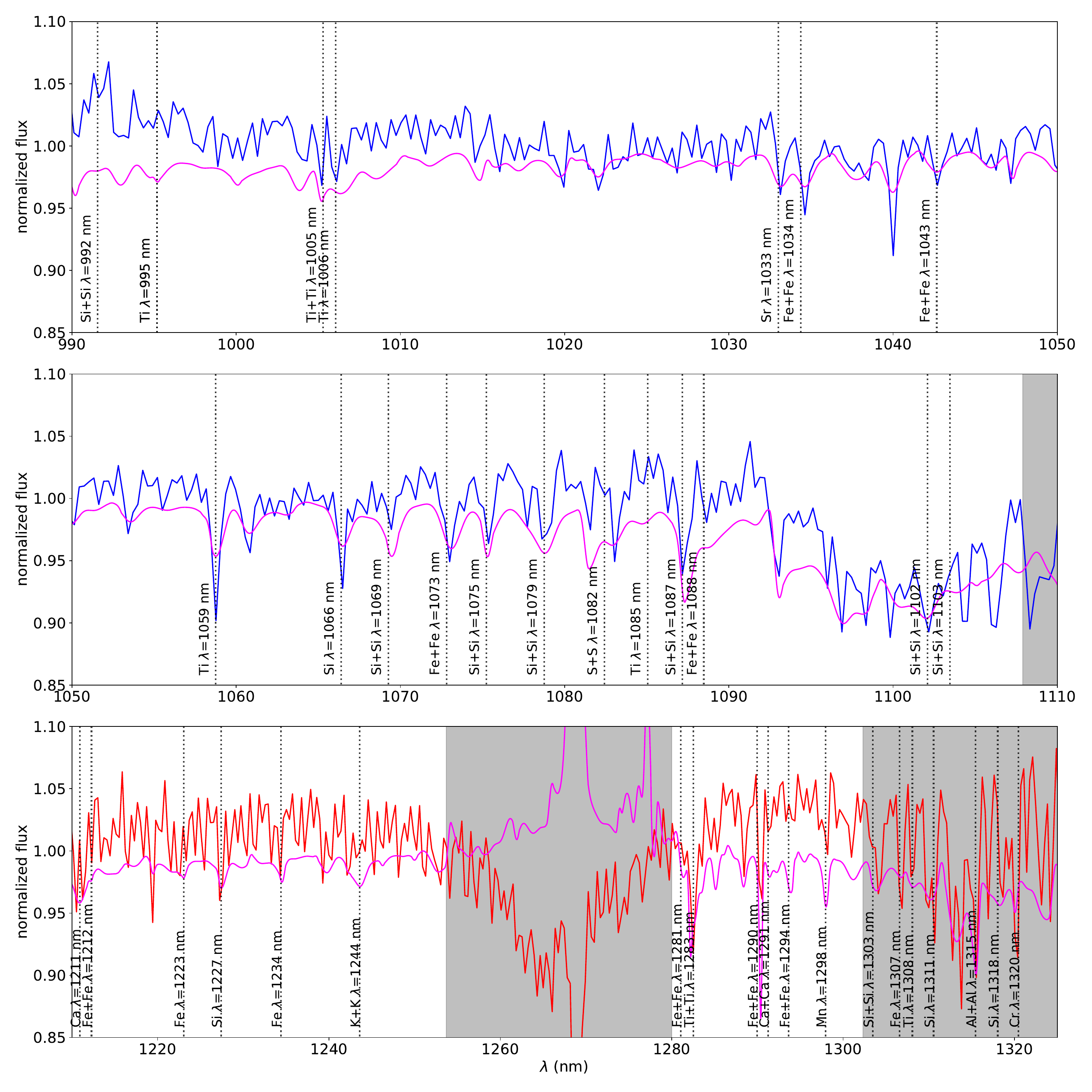}
    \caption{Spectrum of $Arcturus$ from \ac{Lili} (blue and red showing the two orders) compared to a reference spectrum of $Arcturus$ from  \protect\cite{Arcturus_Atlas}, convolved with an approximation of the Lili line spread function (pink). Line identifications are shown with the vertical dotted lines, labeled with the species and wavelength, from \protect\cite{Arcturus_Atlas}. Although there are still imperfections in the continuum normalization of the \ac{Lili} spectra, even subtle features in these spectra match real lines in the reference spectrum.}
    \label{fig:Lili_Arcturus}
\end{figure*}

We also compared the spectrum of \emph{Arcturus} to a reference spectrum from \cite{Arcturus_Atlas} (Fig.~\ref{fig:Lili_Arcturus}). This spectrum was obtained with the Fourier transform spectrometer at Kitt Peak, at a resolution considerably higher than what \ac{Lili} delivers; we therefore convolved this spectrum with a Gaussian function approximating the \ac{Lili} line spread function for a better comparison with our data.  We find the data to have a signal to noise of \qty{\approx 20}{}, due to read noise, dark current and the integration time, which matches our estimations well. Despite this the results of the convolution suggest that we have detected many actual lines in the spectrum and the degree of systematic noise is low.

\subsection{ADC testing}

\begin{figure}
    \centering
    \includegraphics[width=\columnwidth]{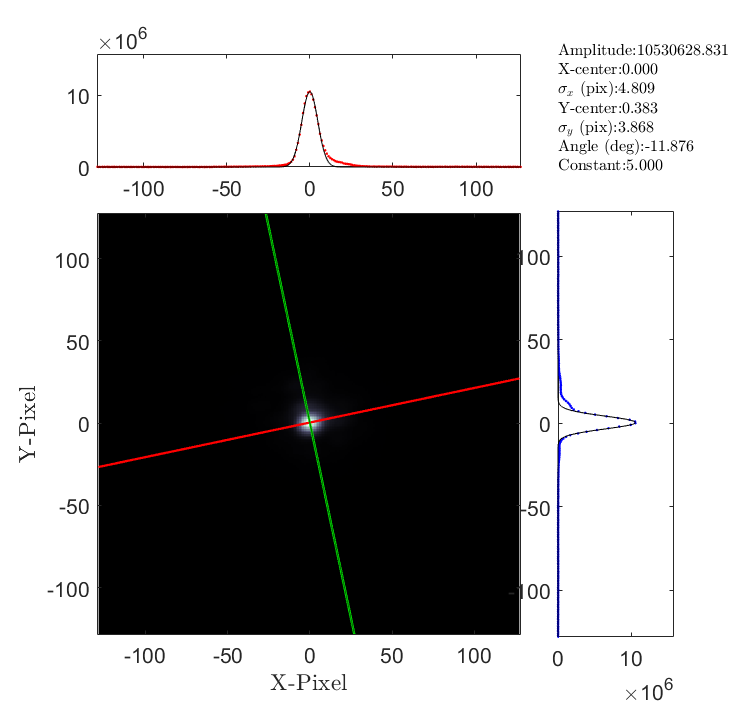}
\includegraphics[width=\columnwidth]{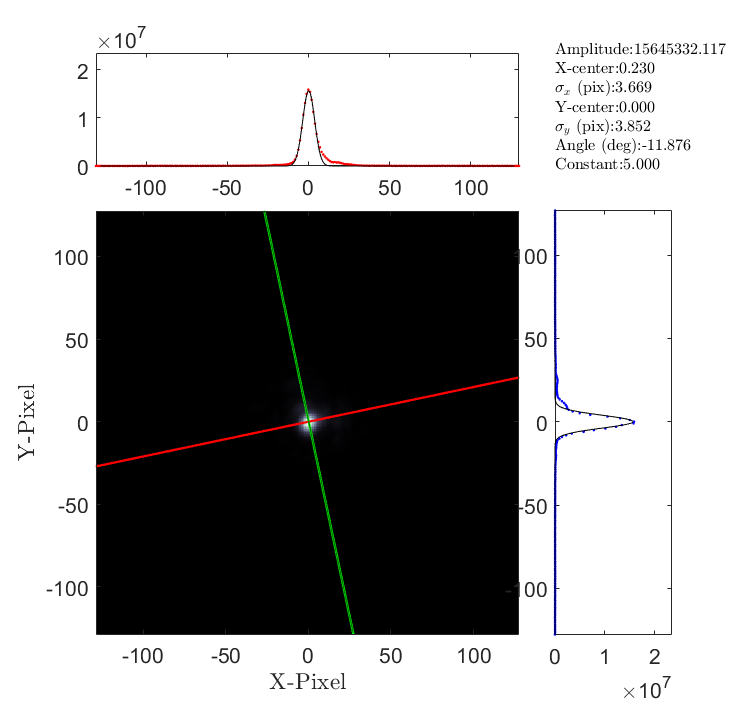}
    \caption{Images  of $27Her$, recorded in the iLocater imaging channel. The images have two dimensional Gaussians fitted, with the $x$ axis set to be along the atmospheric dispersion direction. {\bf upper)} Image of $27Her$ with the \ac{ADC} set to null position. The calculated standard deviation for the Gaussian is 4.8 pixels in the $x$ direction compared to 3.9 in the $y$ direction. {\bf lower)} Image of $27Her$ with the \ac{ADC} on. The calculated standard deviation for the Gaussian is 3.7 pixels in the $x$ direction and 3.9 in the $y$ direction. }
    \label{fig:Lili_27her_imaging}
\end{figure}

\begin{figure*}
    \centering
    \begin{tabular}{cccc}     \includegraphics[width=.25\textwidth]{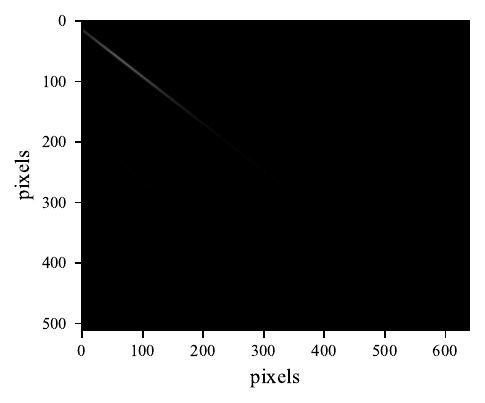}
    \includegraphics[width=.25\textwidth]{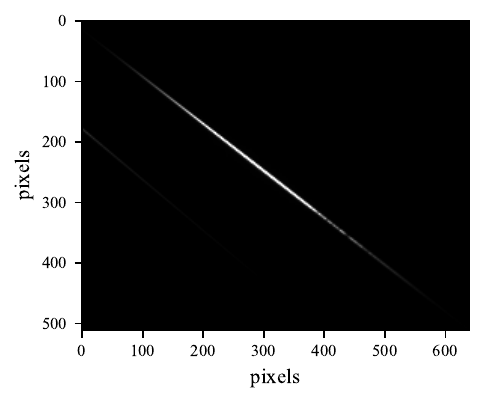}
    \includegraphics[width=.25\textwidth]{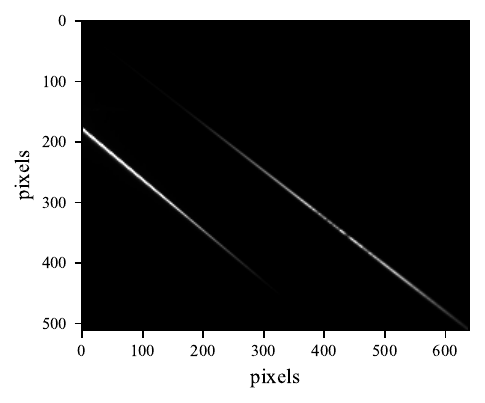}
    \includegraphics[width=.25\textwidth]{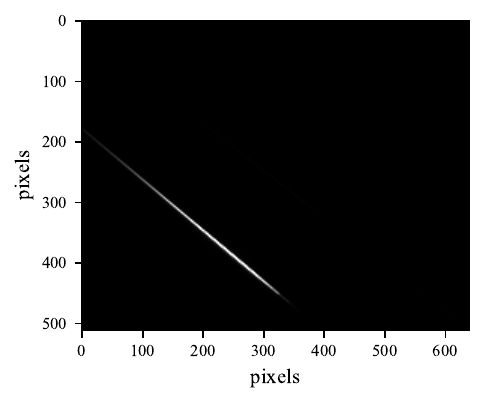}\\
    \includegraphics[width=.25\textwidth]{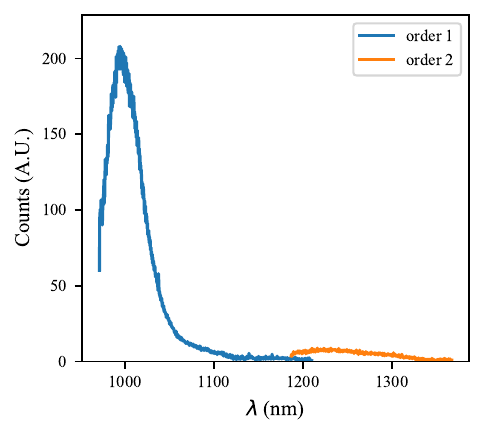}
    \includegraphics[width=.25\textwidth]{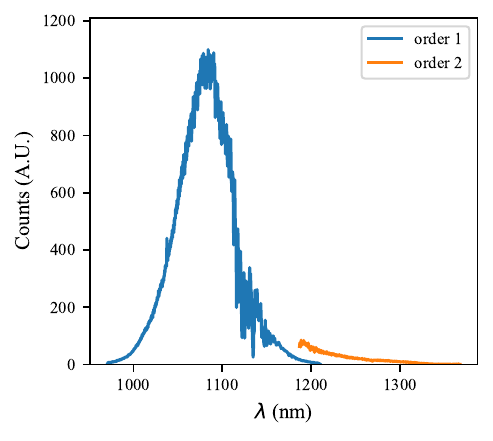}
    \includegraphics[width=.25\textwidth]{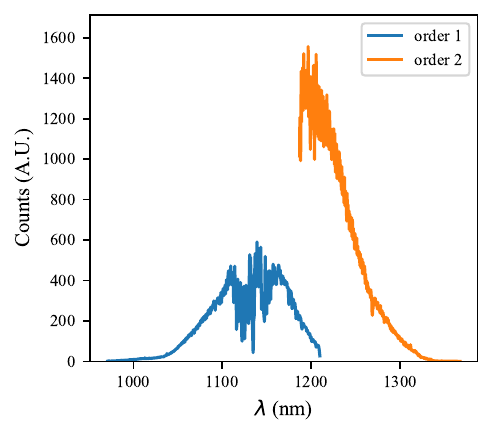}
    \includegraphics[width=.25\textwidth]{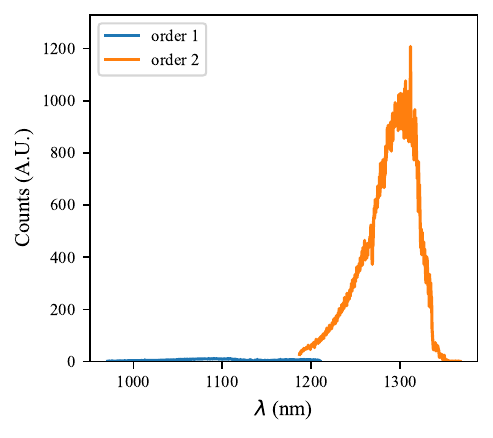}
    
    \end{tabular}
    \caption{Raw frames(top) and extracted spectra (bottom) of $27 Her$ from \ac{Lili}. The four images are of spectra taken with the \ac{ADC} in null position, with the \ac{SMF} being scanned along the atmospheric dispersion direction. As the fibre is scanned, the peak flux changes from the shortest wavelengths (top left in image, left in graph) to the longest (bottom right in image, right in graph).}
    \label{fig:Lili_27her}
\end{figure*}

\begin{figure}
    \centering
    \begin{tabular}{c} 
    \includegraphics[width=\columnwidth]{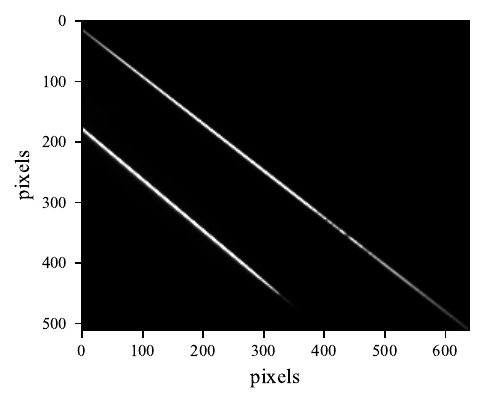}\\
    \includegraphics[width=\columnwidth]{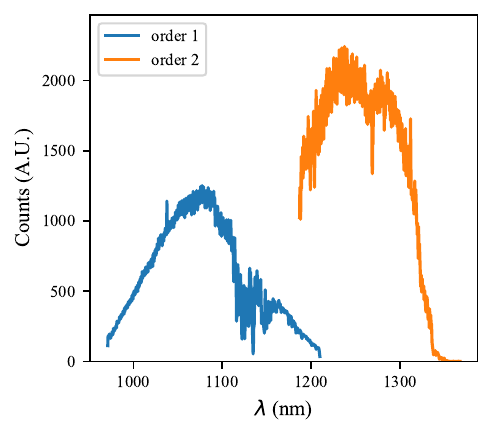}
    \end{tabular}
    \caption{Raw frames and extracted spectra of 27 Her, with the \ac{ADC} set to compensate for chromatic dispersion due to elevation. Here the blaze function of the spectrograph is clearly visible in both orders, but there is flux across the whole iLocater bandpass.}
    \label{fig:Lili_27her_adc}
\end{figure}

Previous testing of the \ac{ADC} within the iLocater acquisition camera \citep{Kopon_2013} had only been completed using the imaging channel \citep{Crass:2021}. \ac{Lili} allowed this to be expanded to detailed characterization across the full iLocater bandpass. The target \emph{27 Her} was used for testing which has a $J_{mag}$ $\approx$ one. While this is a known spectroscopic binary system, its semi major axis of \qty{11}{mas} is below the spatial resolution limit of iLocater and the \ac{LBT}. During observations, \emph{27 Her} was at an elevation of \qty{\approx 40}{\degree} allowing measurable dispersion to be present in both the imaging and spectroscopic channels. Initially the \ac{ADC} was set to its null position (no correction), leading to a dispersed  \ac{PSF} within the acquisition camera (see Fig. \ref{fig:Lili_27her_imaging}, upper). To ensure the dispersion was due to the lack of \ac{ADC} correction, we optimized fibre coupling at the shortest wavelength ($\lambda$ = \qty{0.97}{\micro\meter} and then the iLocater \ac{SMF} was physically scanned across the focal plane in the atmospheric dispersion direction in a series of steps. Moving the \ac{SMF} by \qty{31}{\micro\meter} moved the peaked flux from the blue to the red end of the spectrum ($\lambda$ = \qty{1.31}{\micro\meter}). The results of the stepping procedure on the spectra are shown in Fig. \ref{fig:Lili_27her}. Subsequently, \ac{ADC} correction was enabled within the acquisition camera (see  Fig. \ref{fig:Lili_27her_imaging}, lower) and the expected full \emph{27 Her} spectra was recovered by \ac{Lili} (see Fig. \ref{fig:Lili_27her_adc}).

\subsection{Spatially resolved close companions}

\begin{figure}
    \centering
    \begin{tabular}{c} \includegraphics[width=\columnwidth]{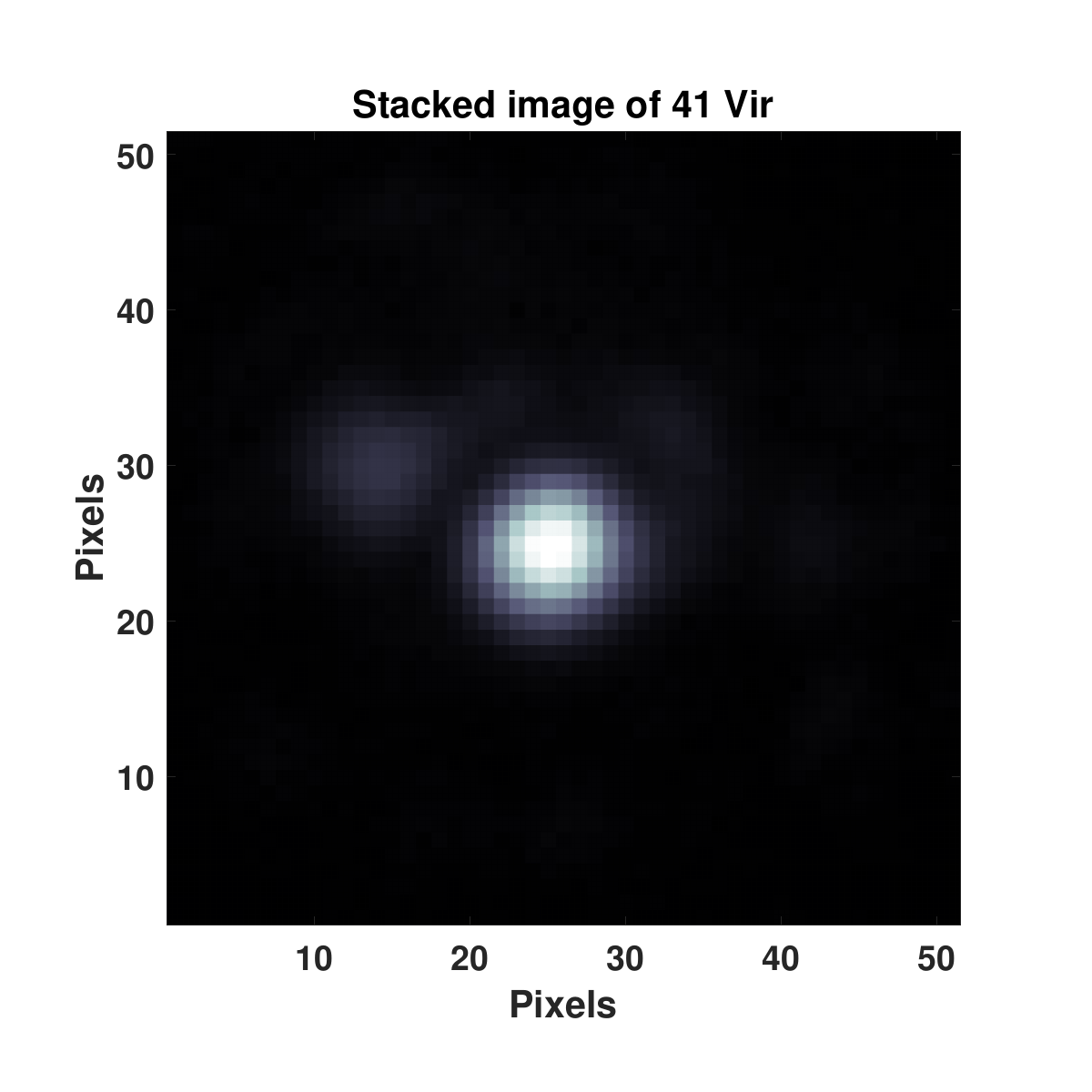}
    \end{tabular}
    \caption{Image of the $41Vir$ system taken with the imaging arm of the iLocater acquisition camera. Here the primary component is clearly visible, with the secondary much fainter above and to the left. Note the distance between the two components is \qty{\approx 50}{mas}.}
\label{fig:Lili_41vir_image}
\end{figure}

\begin{figure*}
    \centering
    \begin{tabular}{c} 
    \includegraphics[width=\textwidth]{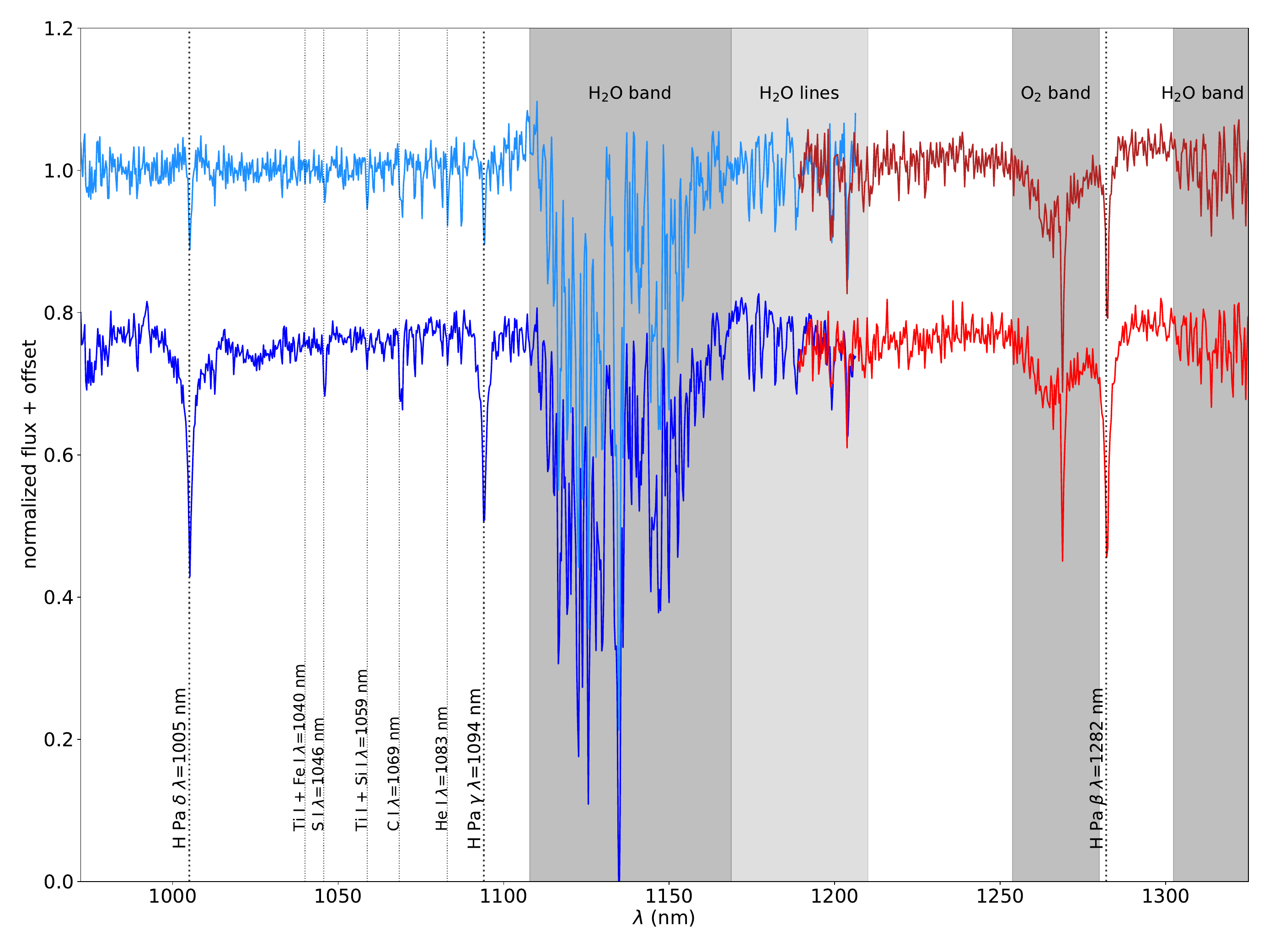}
    \end{tabular}
    \caption{Spectra of the primary component of 41 Vir (bottom) and the 50 mas companion (top), vertically offset for clarity. The two Lili orders are plotted in blue and red respectively. Notable spectral lines and bands are marked.}
    \label{fig:Lili_41vir}
\end{figure*}

The system \emph{41 Vir} (HD 112097) was observed as it is a known double-line spectroscopic binary system, with radial velocity measurements having been made previously \citep{Grandjean2021}. The separation of the two companion stars (\qty{\sim 50}{mas}) means it is difficult to acquire separate spectra of the individual components and the authors have not found any such spectra in the literature. Our observations thus focussed on producing the first spatially resolved spectra of the system. With the telescope \ac{AO} loop closed, the acquisition camera was used to position the \ac{SMF} onto the primary component of the system and spectra were recorded using \ac{Lili}. The fibre was then repositioned onto the second component with spectra again being recorded (see \ref{fig:Lili_41vir_image}). The resulting spectra are shown in Fig. \ref{fig:Lili_41vir}, with the two 41 Vir components  being separated by an offset. The differences in the H Pa $\delta$ and $\gamma$ are clearly visible in the plot and we appear to see more metal lines in the \qtyrange{1050}{1100}{\nano\meter} region in the cooler star, showing that we are observing two different stars.

\section{Future optimisation of Lili}
\label{sec:discussion}

The optical and mechanical components of \ac{Lili} functioned as expected, providing a resolving power of $R \approx 1500$ for the iLocater bandpass. 
Despite this, as discussed in Section \ref{sec:optical_design} and shown in Fig. \ref{fig:Lili_footprint}, \ac{Lili} does not make optimal use of the \ac{CRED2} detection area, nor utilise the full resolving power of the grating. 
It is unlikely that \ac{Lili} will be used for further validation of the iLocater acquisition camera before commissioning in the current timeline. However the design for \ac{Lili} provides a good foundation for future experiments, so improvements are worthwhile considering.

\subsection{Changes to the detector}

To allow optimal sampling while retaining the same \ac{CRED2} detector a similiar approach to the \ac{MCIFU} could be taken \citep{Haffert2020}. This would involve removing the second lens, increasing the size of the image and moving the \ac{CRED2} using two linear stages to cover the whole bandpass. For the tests in this paper, this was not an option as we wished to cover the whole iLocater bandpass simultaneously, but for other experiments could provide the whole bandpass at a higher resolving power. Alternatively a detector with more pixels (such as a HAWAII-2) could be purchased, allowing the whole field to be covered simultaneously. This would no longer allow the same frame rate as the \ac{CRED2}, but would have lower noise and higher \ac{QE}.

\subsection{Image Slicing to optimise detector use}

To efficiently fit the spectrum from \ac{Lili} onto the \ac{CRED2} detector an image slicer could be used. Image slicers are typically used as a \acp{IFU} \citep{Calcines2013} to reorganise a bi-dimensional field of view into one or more slits that illuminate a spectrograph, thereby obtaining the spectra of all spatial elements simultaneously. Image slicers can also be found at a pupil position \citep{Roelfsema2018} (pupil slicer) or as a field reformatter \citep{Calcines2023} before the spectrograph. We investigated using an image slicer to slice the output spectrum and image it on the detector. The investigation revealed a number of problems with this approach and it was decided this would be too costly and difficult to implement. In the future we plan to investigate alternative solutions, like the use of an array of slicer mirrors to decompose the spectrum and correct its tilt angle, combined with lenses to fit the spectra on the detector. These lenses will be optimized for their correspondent wavelengths, reducing the volume required for spectrum slicing.

\section{Conclusions}\label{sec:conclusion}\acresetall

The results presented in this paper show the first spectral characterisation of the iLocater acquisition camera at the \ac{LBT}. To achieve them we designed, built and tested \ac{Lili}, a \ac{SMF}-fed spectrograph covering the iLocater wavelength range \qtyrange{0.97}{1.31}{\micro\meter} with a resolving power between \qtyrange{1500}{2100}{}. The resolving power and tranmission of the as-built spectrograph matched the design.

The tests detailed in this paper and previously \citep{Crass:2021} are key to  ensuring an efficient commissioning process for iLocater.
By studying bright stars at high frame rates, we were able to record and extract spectra across the full iLocater passband which is consistent with prior data in the literature. Our \ac{ADC} tests show good performance at elevations of \qty{40}{\degree}, which is the approaching the operational limit for the \ac{ADC} and telescope \ac{AO} system \citep{Crass:2021}. This complements and extends the confidence that during during commissioning, that iLocater will be able to push to the lower limit elevation of \qty{30}{\degree}. We believe our spectra are the first spatially resolved spectra of $41Vir$ A and B. In acquiring them, we have demonstrated the ability of the acquisition camera to spectrally separate close-separation systems (\qty{\approx 50}{mas}) and such a proven capability will help define early science targets for iLocater and subsequent studies.

As this and previous papers have shown, the facilities at the \ac{LBT} and the iLocater acquisition camera provide an excellent platform to test new technologies and techniques  \citep{Hottinger2021}. Future testing with this system include the commisioning of iLocater and the full optimization of the acquisition camera. As this platform also enables the demonstration of photonic technologies in astronomical instrumentation, such as photonic beam stabilization technologies, photonic spectrographs or novel types of \acp{IFU}, it can play an important role in their timely advancement. This will help augment diffraction-limited on-sky capabilities in an efficient way, and deliver science capabilities on an accelerated timeline.

\subsection*{Acknowledgments}\label{sec:acknowledgements}

R.J.H would like to thank all that have indirectly contributed to the project. In particular André Boné for his optical design input and the MCIFU team for their work designing the \ac{MCIFU}, inspiring this spectrograph. He wants to thank Mark Swinbank and Daniel J. Stevens for  discussions on astronomical targets and spectra. He also wishes to thank Lazar Staykov, the support team at the \ac{LBT} and the L3 lab technicians Durham Uni, without whom this would not have worked out. 

M.C.J. thanks Jennifer Johnson and Adam Wheeler for useful discussions on line identification and spectral synthesis. He also acknowledges support from the Thomas Jefferson Endowment for Space Exploration.

The LBT is an international collaboration among institutions in the United States, Italy and Germany. LBT Corporation Members are: The University of Arizona on behalf of the Arizona Board of Regents; Istituto Nazionale di Astrofisica, Italy; LBT Beteiligungsgesellschaft, Germany, representing the Max-Planck Society, The Leibniz Institute for Astrophysics Potsdam, and Heidelberg University; The Ohio State University, representing OSU, University of Notre Dame, University of Minnesota and University of Virginia. 

 Observations have benefited from the use of ALTA Center (alta.arcetri.inaf.it) forecasts performed with the Astro-Meso-Nh model. Initialization data of the ALTA automatic forecast system come from the General Circulation Model (HRES) of the European Centre for Medium Range Weather Forecasts.

This material is based upon work supported by the National Science Foundation under Grant No. 1654125 \& 2108603.

This research made use of Astropy, a community-developed core \texttt{Python} package for \cite{astropy:2013, astropy:2018}, Numpy \cite{numpy},  Scipy \cite{2020SciPy-NMeth} and Matplotlib \cite{matplotlib}.

\subsection*{Data Availability}
The data and spectroscopic reduction pipeline underlying this article will be shared on reasonable request to the corresponding author.


\bibliographystyle{mnras}
\bibliography{report} 

\begin{thebibliography}{}
\makeatletter
\relax
\def\mn@urlcharsother{\let\do\@makeother \do\$\do\&\do\#\do\^\do\_\do\%\do\~}
\def\mn@doi{\begingroup\mn@urlcharsother \@ifnextchar [ {\mn@doi@}
  {\mn@doi@[]}}
\def\mn@doi@[#1]#2{\def\@tempa{#1}\ifx\@tempa\@empty \href
  {http://dx.doi.org/#2} {doi:#2}\else \href {http://dx.doi.org/#2} {#1}\fi
  \endgroup}
\def\mn@eprint#1#2{\mn@eprint@#1:#2::\@nil}
\def\mn@eprint@arXiv#1{\href {http://arxiv.org/abs/#1} {{\tt arXiv:#1}}}
\def\mn@eprint@dblp#1{\href {http://dblp.uni-trier.de/rec/bibtex/#1.xml}
  {dblp:#1}}
\def\mn@eprint@#1:#2:#3:#4\@nil{\def\@tempa {#1}\def\@tempb {#2}\def\@tempc
  {#3}\ifx \@tempc \@empty \let \@tempc \@tempb \let \@tempb \@tempa \fi \ifx
  \@tempb \@empty \def\@tempb {arXiv}\fi \@ifundefined
  {mn@eprint@\@tempb}{\@tempb:\@tempc}{\expandafter \expandafter \csname
  mn@eprint@\@tempb\endcsname \expandafter{\@tempc}}}

\bibitem[\protect\citeauthoryear{Anagnos et~al.,}{Anagnos
  et~al.}{2018}]{anagnos2018}
Anagnos T.,  et~al., 2018, \mn@doi [Monthly Notices of the Royal Astronomical
  Society] {10.1093/mnras/sty1396}, 478, 4881

\bibitem[\protect\citeauthoryear{{Astropy Collaboration} et~al.,}{{Astropy
  Collaboration} et~al.}{2013}]{astropy:2013}
{Astropy Collaboration} et~al., 2013, \mn@doi [\aap]
  {10.1051/0004-6361/201322068}, \href
  {http://adsabs.harvard.edu/abs/2013A%26A...558A..33A} {558, A33}

\bibitem[\protect\citeauthoryear{{Bechter}, {Bechter}, {Crepp}, {King}  \&
  {Crass}}{{Bechter} et~al.}{2018}]{bechter_18}
{Bechter} A.~J.,  {Bechter} E.~B.,  {Crepp} J.~R.,  {King} D.,   {Crass} J.,
  2018, in {Evans} C.~J.,  {Simard} L.,   {Takami} H.,  eds,  Society of
  Photo-Optical Instrumentation Engineers (SPIE) Conference Series Vol. 10702,
  Ground-based and Airborne Instrumentation for Astronomy VII. p. 107026T
  (\mn@eprint {arXiv} {1812.02704}), \mn@doi{10.1117/12.2313658}

\bibitem[\protect\citeauthoryear{{Bechter}, {Bechter}, {Crepp}  \&
  {Crass}}{{Bechter} et~al.}{2019}]{bechter_19_hr4g}
{Bechter} E.~B.,  {Bechter} A.~J.,  {Crepp} J.~R.,   {Crass} J.,  2019, \mn@doi
  [Journal of Astronomical Telescopes, Instruments, and Systems]
  {10.1117/1.JATIS.5.3.038004}, \href
  {https://ui.adsabs.harvard.edu/abs/2019JATIS...5c8004B} {5, 038004}

\bibitem[\protect\citeauthoryear{{Bechter}, {Bechter}, {Crepp}, {Ketterer}  \&
  {Crass}}{{Bechter} et~al.}{2020}]{bechter_20_pol}
{Bechter} A.~J.,  {Bechter} E.~B.,  {Crepp} J.~R.,  {Ketterer} R.,   {Crass}
  J.,  2020, \mn@doi [\pasp] {10.1088/1538-3873/ab9cc6}, \href
  {https://ui.adsabs.harvard.edu/abs/2020PASP..132i5001B} {132, 095001}

\bibitem[\protect\citeauthoryear{Birks, Gris-S{\'a}nchez, Yerolatsitis,
  Leon-Saval  \& Thomson}{Birks et~al.}{2015}]{Birks:2015}
Birks T.~A.,  Gris-S{\'a}nchez I.,  Yerolatsitis S.,  Leon-Saval S.,   Thomson
  R.~R.,  2015, Advances in Optics and Photonics, 7, 107

\bibitem[\protect\citeauthoryear{Bland-Hawthorn \& Horton}{Bland-Hawthorn \&
  Horton}{2006}]{Bland-Hawthorn2006}
Bland-Hawthorn J.,  Horton A.,  2006, in McLean I.~S.,  Iye M.,  eds,  Society
  of Photo-Optical Instrumentation Engineers (SPIE) Conference Series Vol.
  6269, Ground-based and Airborne Instrumentation for Astronomy. SPIE, p.
  62690N, \mn@doi{10.1117/12.670931}, \url {https://doi.org/10.1117/12.670931}

\bibitem[\protect\citeauthoryear{Bohlin, Gordon  \& Tremblay}{Bohlin
  et~al.}{2014}]{Bohlin_2014}
Bohlin R.~C.,  Gordon K.~D.,   Tremblay P.-E.,  2014, \mn@doi [Publications of
  the Astronomical Society of the Pacific] {10.1086/677655}, 126, 711

\bibitem[\protect\citeauthoryear{Calcines, L\'{o}pez  \& Collados}{Calcines
  et~al.}{2013}]{Calcines2013}
Calcines A.,  L\'{o}pez R.~L.,   Collados M.,  2013, \mn@doi [Journal of
  Astronomical Instrumentation] {10.1142/S2251171713500098}, 02, 1350009

\bibitem[\protect\citeauthoryear{{Calcines}, {Wells}, {O'Brien}, {Morris},
  {Seifert}, {Zanutta}, {Evans}  \& {Di Marcantonio}}{{Calcines}
  et~al.}{2023}]{Calcines2023}
{Calcines} A.,  {Wells} M.,  {O'Brien} K.,  {Morris} S.,  {Seifert} W.,
  {Zanutta} A.,  {Evans} C.,   {Di Marcantonio} P.,  2023, \mn@doi
  [Experimental Astronomy] {10.1007/s10686-022-09866-5}, \href
  {https://ui.adsabs.harvard.edu/abs/2023ExA....55..267C} {55, 267}

\bibitem[\protect\citeauthoryear{{Crass} et~al.,}{{Crass}
  et~al.}{2019}]{Crass2019}
{Crass} J.,  et~al., 2019, in Bulletin of the American Astronomical Society.
  p.~122

\bibitem[\protect\citeauthoryear{Crass et~al.,}{Crass
  et~al.}{2021}]{Crass:2021}
Crass J.,  et~al., 2021, \mn@doi [Monthly Notices of the Royal Astronomical
  Society] {10.1093/mnras/staa3355}, 501, 2250

\bibitem[\protect\citeauthoryear{{Crass} et~al.,}{{Crass}
  et~al.}{2022}]{Crass_iLocSpec:2021}
{Crass} J.,  et~al., 2022, in {Evans} C.~J.,  {Bryant} J.~J.,   {Motohara} K.,
  eds,  Society of Photo-Optical Instrumentation Engineers (SPIE) Conference
  Series Vol. 12184, Ground-based and Airborne Instrumentation for Astronomy
  IX. p. 121841P (\mn@eprint {arXiv} {2209.00009}), \mn@doi{10.1117/12.2630228}

\bibitem[\protect\citeauthoryear{{Crepp}}{{Crepp}}{2014}]{Crepp2014}
{Crepp} J.~R.,  2014, \mn@doi [Science] {10.1126/science.1262071}, \href
  {https://ui.adsabs.harvard.edu/abs/2014Sci...346..809C} {346, 809}

\bibitem[\protect\citeauthoryear{Crepp et~al.,}{Crepp et~al.}{2016}]{Crepp2016}
Crepp J.~R.,  et~al., 2016, in Ground-based and Airborne Instrumentation for
  Astronomy VI. p. 990819

\bibitem[\protect\citeauthoryear{{Ertel} et~al.,}{{Ertel}
  et~al.}{2020}]{Ertel2020}
{Ertel} S.,  et~al., 2020, in {Tuthill} P.~G.,  {M{\'e}rand} A.,   {Sallum} S.,
   eds,  Society of Photo-Optical Instrumentation Engineers (SPIE) Conference
  Series Vol. 11446, Optical and Infrared Interferometry and Imaging VII. p.
  1144607, \mn@doi{10.1117/12.2561849}

\bibitem[\protect\citeauthoryear{{GRAVITY Collaboration} et~al.,}{{GRAVITY
  Collaboration} et~al.}{2017}]{gravity:2017}
{GRAVITY Collaboration} et~al., 2017, \mn@doi [A\&A]
  {10.1051/0004-6361/201730838}, 602, A94

\bibitem[\protect\citeauthoryear{Gibson, Oppenheimer, Matthews  \&
  Vasisht}{Gibson et~al.}{2019}]{Gibson2019}
Gibson R.~K.,  Oppenheimer R.,  Matthews C.~T.,   Vasisht G.,  2019, \mn@doi
  [Journal of Astronomical Telescopes, Instruments, and Systems]
  {10.1117/1.JATIS.6.1.011002}, 6, 011002

\bibitem[\protect\citeauthoryear{{Grandjean, A.} et~al.,}{{Grandjean, A.}
  et~al.}{2021}]{Grandjean2021}
{Grandjean, A.} et~al., 2021, \mn@doi [A&A] {10.1051/0004-6361/202039672}, 650,
  A39

\bibitem[\protect\citeauthoryear{Haffert et~al.,}{Haffert
  et~al.}{2020}]{Haffert2020}
Haffert S.~Y.,  et~al., 2020, \mn@doi [Journal of Astronomical Telescopes,
  Instruments, and Systems] {10.1117/1.JATIS.6.4.045007}, 6, 045007

\bibitem[\protect\citeauthoryear{Harris \& Allington-Smith}{Harris \&
  Allington-Smith}{2012}]{Harris2012}
Harris R.~J.,  Allington-Smith J.~R.,  2012, \mn@doi [Monthly Notices of the
  Royal Astronomical Society] {10.1093/mnras/sts265}, 428, 3139

\bibitem[\protect\citeauthoryear{Harris et~al.,}{Harris
  et~al.}{2015}]{Harris:2015}
Harris R.~J.,  et~al., 2015, \mn@doi [\mnras] {10.1093/mnras/stv410}, 450, 428

\bibitem[\protect\citeauthoryear{Harris et~al.,}{Harris et~al.}{2020}]{numpy}
Harris C.~R.,  et~al., 2020, \mn@doi [Nature] {10.1038/s41586-020-2649-2}, 585,
  357–362

\bibitem[\protect\citeauthoryear{{Hinkle}, {Wallace}  \& {Livingston}}{{Hinkle}
  et~al.}{1995}]{Arcturus_Atlas}
{Hinkle} K.,  {Wallace} L.,   {Livingston} W.,  1995, \mn@doi [\pasp]
  {10.1086/133660}, \href
  {https://ui.adsabs.harvard.edu/abs/1995PASP..107.1042H} {107, 1042}

\bibitem[\protect\citeauthoryear{{Hinz} et~al.,}{{Hinz}
  et~al.}{2016}]{Hinz2016}
{Hinz} P.~M.,  et~al., 2016, in {Malbet} F.,  {Creech-Eakman} M.~J.,
  {Tuthill} P.~G.,  eds,  Society of Photo-Optical Instrumentation Engineers
  (SPIE) Conference Series Vol. 9907, Optical and Infrared Interferometry and
  Imaging V. p. 990704, \mn@doi{10.1117/12.2233795}

\bibitem[\protect\citeauthoryear{{Horne}}{{Horne}}{1986}]{Horne:1986}
{Horne} K.,  1986, \mn@doi [\pasp] {10.1086/131801}, \href
  {https://ui.adsabs.harvard.edu/abs/1986PASP...98..609H} {98, 609}

\bibitem[\protect\citeauthoryear{Hottinger et~al.,}{Hottinger
  et~al.}{2021}]{Hottinger2021}
Hottinger P.,  et~al., 2021, \mn@doi [J. Opt. Soc. Am. B]
  {10.1364/JOSAB.421459}, 38, 2517

\bibitem[\protect\citeauthoryear{{Hunter}}{{Hunter}}{2007}]{matplotlib}
{Hunter} J.~D.,  2007, \mn@doi [Computing in Science and Engineering]
  {10.1109/MCSE.2007.55}, \href
  {https://ui.adsabs.harvard.edu/abs/2007CSE.....9...90H} {9, 90}

\bibitem[\protect\citeauthoryear{Jones, Noll, Kausch, Szyszka  \&
  Kimeswenger}{Jones et~al.}{2013}]{Jones2013}
Jones A.,  Noll S.,  Kausch W.,  Szyszka C.,   Kimeswenger S.,  2013, \mn@doi
  [A&A] {10.1051/0004-6361/201322433}, 560, A91

\bibitem[\protect\citeauthoryear{Jovanovic et~al.,}{Jovanovic
  et~al.}{2017}]{Jovanovic2017}
Jovanovic N.,  et~al., 2017, \mn@doi [A&A] {10.1051/0004-6361/201630351}, 604,
  A122

\bibitem[\protect\citeauthoryear{Kervella et~al.,}{Kervella
  et~al.}{2003}]{Kervella2003}
Kervella P.,  et~al., 2003, in Traub W.~A.,  ed., ~SPIE Vol. 4838,
  Interferometry for Optical Astronomy II. SPIE, pp 858 -- 869,
  \mn@doi{10.1117/12.459345}, \url {https://doi.org/10.1117/12.459345}

\bibitem[\protect\citeauthoryear{Kopon, Close, Males  \& Gasho}{Kopon
  et~al.}{2013}]{Kopon_2013}
Kopon D.,  Close L.~M.,  Males J.~R.,   Gasho V.,  2013, \mn@doi [Publications
  of the Astronomical Society of the Pacific] {10.1086/672091}, 125, 966–975

\bibitem[\protect\citeauthoryear{{Le Bouquin, J.-B.} et~al.,}{{Le Bouquin,
  J.-B.} et~al.}{2011}]{LeBouquin2011}
{Le Bouquin, J.-B.} et~al., 2011, \mn@doi [A&A] {10.1051/0004-6361/201117586},
  535, A67

\bibitem[\protect\citeauthoryear{Leon-Saval, Birks, Bland-Hawthorn  \&
  Englund}{Leon-Saval et~al.}{2005}]{Leon-Saval:2005}
Leon-Saval S.,  Birks T.,  Bland-Hawthorn J.,   Englund M.,  2005, Optics
  letters, 30, 2545

\bibitem[\protect\citeauthoryear{Lovis et~al.,}{Lovis et~al.}{2022}]{Lovis2022}
Lovis C.,  et~al., 2022, in Evans C.~J.,  Bryant J.~J.,   Motohara K.,  eds,
  ~SPIE Vol. 12184, Ground-based and Airborne Instrumentation for Astronomy IX.
  SPIE, p. 121841Q, \mn@doi{10.1117/12.2627923}, \url
  {https://doi.org/10.1117/12.2627923}

\bibitem[\protect\citeauthoryear{{MacLachlan} et~al.,}{{MacLachlan}
  et~al.}{2017}]{MacLachlan:2017}
{MacLachlan} D.~G.,  et~al., 2017, \mn@doi [\mnras] {10.1093/mnras/stw2558},
  \href {http://adsabs.harvard.edu/abs/2017MNRAS.464.4950M} {464, 4950}

\bibitem[\protect\citeauthoryear{Mawet et~al.,}{Mawet et~al.}{2022}]{Mawet2022}
Mawet D.,  et~al., 2022, in Evans C.~J.,  Bryant J.~J.,   Motohara K.,  eds,
  ~SPIE Vol. 12184, Ground-based and Airborne Instrumentation for Astronomy IX.
  SPIE, p. 121841R, \mn@doi{10.1117/12.2630142}, \url
  {https://doi.org/10.1117/12.2630142}

\bibitem[\protect\citeauthoryear{{Moehler, S.} et~al.,}{{Moehler, S.}
  et~al.}{2014}]{Moehler2014}
{Moehler, S.} et~al., 2014, \mn@doi [A&A] {10.1051/0004-6361/201423790}, 568,
  A9

\bibitem[\protect\citeauthoryear{{Noll, S.}, {Kausch, W.}, {Barden, M.},
  {Jones, A. M.}, {Szyszka, C.}, {Kimeswenger, S.}  \& {Vinther, J.}}{{Noll,
  S.} et~al.}{2012}]{Noll2012}
{Noll, S.} {Kausch, W.} {Barden, M.} {Jones, A. M.} {Szyszka, C.} {Kimeswenger,
  S.}  {Vinther, J.} 2012, \mn@doi [A&A] {10.1051/0004-6361/201219040}, 543,
  A92

\bibitem[\protect\citeauthoryear{Pinna et~al.,}{Pinna et~al.}{2016}]{Pinna2016}
Pinna E.,  et~al., 2016, in Marchetti E.,  Close L.~M.,   V{\'e}ran J.-P.,
  eds, ~SPIE Vol. 9909, Adaptive Optics Systems V. SPIE, p. 99093V,
  \mn@doi{10.1117/12.2234444}, \url {https://doi.org/10.1117/12.2234444}

\bibitem[\protect\citeauthoryear{{Price-Whelan} et~al.,}{{Price-Whelan}
  et~al.}{2018}]{astropy:2018}
{Price-Whelan} A.~M.,  et~al., 2018, \mn@doi [\aj] {10.3847/1538-3881/aabc4f},
  \href {https://ui.adsabs.harvard.edu/#abs/2018AJ....156..123T} {156, 123}

\bibitem[\protect\citeauthoryear{Roelfsema et~al.,}{Roelfsema
  et~al.}{2018}]{Roelfsema2018}
Roelfsema R.,  et~al., 2018, in Evans C.~J.,  Simard L.,   Takami H.,  eds,
  ~SPIE Vol. 10702, Ground-based and Airborne Instrumentation for Astronomy
  VII. SPIE, p. 107027V, \mn@doi{10.1117/12.2312473}, \url
  {https://doi.org/10.1117/12.2312473}

\bibitem[\protect\citeauthoryear{Scott, Millan-Gabert, Lhom\'{e},
  Ten~Brummelarr, Coud\'{e} Du~Foresto, Sturmann  \& Sturmann}{Scott
  et~al.}{2013}]{Scott:2013}
Scott N.~J.,  Millan-Gabert R.,  Lhom\'{e} E.,  Ten~Brummelarr T.~A.,
  Coud\'{e} Du~Foresto V.,  Sturmann J.,   Sturmann L.,  2013, \mn@doi [Journal
  of Astronomical Instrumentation] {10.1142/S2251171713400059}, 02, 1340005

\bibitem[\protect\citeauthoryear{Sliski, Blake, Eastman  \& Halverson}{Sliski
  et~al.}{2023}]{Sliski_2023}
Sliski D.~H.,  Blake C.~H.,  Eastman J.~D.,   Halverson S.,  2023, \mn@doi
  [Astronomische Nachrichten] {10.1002/asna.20220080}, 344

\bibitem[\protect\citeauthoryear{Spaleniak, Jovanovic, Gross, Ireland, Lawrence
   \& Withford}{Spaleniak et~al.}{2013}]{Spaleniak2013}
Spaleniak I.,  Jovanovic N.,  Gross S.,  Ireland M.~J.,  Lawrence J.~S.,
  Withford M.~J.,  2013, \mn@doi [Opt. Express] {10.1364/OE.21.027197}, 21,
  27197

\bibitem[\protect\citeauthoryear{{Vigan, A.} et~al.,}{{Vigan, A.}
  et~al.}{2024}]{Vigan2024}
{Vigan, A.} et~al., 2024, \mn@doi [A&A] {10.1051/0004-6361/202348019}, 682, A16

\bibitem[\protect\citeauthoryear{Virtanen et~al.,}{Virtanen
  et~al.}{2020}]{2020SciPy-NMeth}
Virtanen P.,  et~al., 2020, \mn@doi [Nature Methods]
  {10.1038/s41592-019-0686-2}, \href {https://rdcu.be/b08Wh} {17, 261}

\makeatother
\end{thebibliography}








\bsp	
\label{lastpage}
\end{document}